\documentclass[prc,twocolumn,showpacs,superscriptaddress]{revtex4}
\usepackage{rotating}
\usepackage{amsmath,amssymb}
\usepackage{bm,subfigure}
\usepackage{verbatim}
\usepackage{graphicx}
\usepackage{dcolumn}
\usepackage{amstext}
\usepackage{lscape}
\usepackage{mathptmx}
\usepackage{verbatim}
\usepackage[colorlinks,citecolor=blue,bookmarks]{hyperref}
\usepackage{times}

\begin{document}

\title{Effects of isovector scalar meson on hyperon star}
 
\author{S. K. Biswal, Bharat Kumar and S. K. Patra}

\affiliation{
Institute of Physics, Bhubaneswar-751005, India}
\date{\today}

\begin{abstract}
We study the effects of  isovector-scalar  ($\delta$)-meson on neutron 
star. Influence of $\delta$-meson on both static and rotating neutron star 
is discussed. Inclusion of $\delta$-meson in a neutron star system 
consisting of proton, neutron and electron, make the equation of state 
stiffer in higher density and consequently increases the maximum mass of 
the star. But induction of $\delta$-meson in the hyperon star 
decreases the maximum mass of the hyperon star. This is due to the early 
evolution of hyperons in presence of $\delta-$meson. 
\end{abstract}
\pacs {   21.10.Dr,  23.40.-s,  23.60.+e,  24.75.+i}
\maketitle
\footnotetext [1]{sbiswal@iopb.res.in}
\footnotetext [1]{bharat@iopb.res.in}
\footnotetext [1]{patra@iopb.res.in}
\section{Introduction}\label{sec1}
Neutron star is a venerable candidate to discuss the physics at 
high density. We can not create such a high density in terrestrial
 laboratory, so 
neutron star is and the only object, which can provides many information on 
high density nature of the matter\cite{prakash04,prakash07}. But it is not an 
easy task to deal with 
the neutron star for it's complex nature, as all the 
four fundamental forces (strong, weak, gravitational and electromagnetic) 
are active.  High gravitational field makes  
mandatory to use general theory of relativity for the study of neutron star 
structure. Equations of states (EOS) are the sole ingredient that must be 
supplied to the equation of stellar structure,
Tolman-Oppenheimer-Volkoff (TOV) equation, whose out-come 
is the mass-radius profile of the dense neutron star. 
In this case, the nuclear EOS plays an intimate 
role in deciding the mass-radius of a neutron star. Its indispensable 
important attracts the attention of physicists to have an anatomy of
the interactions Lagrangian.
As the name suggests, neutron star is not completely made up 
neutrons, a small fraction of protons and electrons  are also present, 
which is the consequence of the $\beta-$equilibrium and charge neutrality 
condition\cite{glen85}. Also, the presence of exotic degrees of freedom like 
hyperons and kaons can not be ignored in such a high dense matter. 
It is one among the most asymmetric and dense nuclear system  
in  nature. 

From last three decades \cite{rein89,ring96}, the relativistic mean field (RMF)
generalized by Walecka \cite{walecka74} and later on developed by Boguta and 
Bodmer \cite{boguta77} is one amongst the most reliable theory 
to deal with  infinite nuclear matter and finite nuclei. The original RMF 
formalism starts with an effective Lagrangian, whose degrees of freedom are
nucleons, $\sigma-,$ $\omega-$, $\rho-$ and $\pi-$mesons. To reproduce proper
experimental observable, it is extended to the self-interaction of 
$\sigma-$meson.  Recently, all other self- and crossed interactions including
the baryon octet are also introduced keeping in view the extra-ordinary 
condition of the system, such as highly asymmetric system or extremely high 
density medium \cite{bharat07}. 
Since the RMF formalism is an effective nucleons-mesons model, the
coupling constants for both nucleon-meson and hyperon-meson are fitted
to reproduce the properties of selected nuclei and infinite nuclear matter
properties \cite{walecka74,boguta77,rein86,ring90}. In this case, it is
improper to use the parameters obtained from the free nucleon-nucleon 
scattering data. 
The parameters, with proper relativistic kinematics and with the
mesons and their properties already known or fixed from the properties of a 
small number of finite nuclei, the method gives excellent results 
not only for spherical nuclei, but also of well-known deformed cases.
The same force parametrization can be used both for $\beta-$stable 
and $\beta-$unstable nuclei through-out the periodic table
\cite{rein88,sharma93,suga94,lala97}. 

The importance of the self- and crossed- interactions  are significant for
some specific properties of nuclei/nuclear-matter in certain conditions. 
For example, self-interaction of  $\sigma$-meson takes care of the reduction
of nuclear matter incompressibility $K_{\infty}$ from an unacceptable high
value of $K_{\infty}\sim 600$ MeV to a reasonable number of $\sim 270$ MeV
 \cite{boguta77,bodmer91}, while the self-interaction 
of vector meson $\omega$ soften the equation of state\cite{suga94,pieka05}. 
Thus, it is imperative to include all the mesons and their possible interactions
with nucleons, self- and crossed terms in the effective Lagrangian density. 
However, it is not necessary to do so, because of the symmetry reason and 
their heavy masses \cite{mechl87}. For example, to keep the spin-isospin 
and parity symmetry in the ground state, the contribution of $\pi-$meson 
is ignored \cite{green72} and also the effect of heaver mesons are neglected for their
negligible contribution. Taking into this argument, in many versions of the
RMF formalisms, the inclusion of isovector-scalar ($\delta$) meson is neglected 
due to its small contribution. 
But  recently it is seen \cite{sing14,shailesh14b,kubis97,roca11}
that the endowment of the $\delta$-meson goes on increasing with  density 
and  asymmetry of  nuclear system. 
Thus, it will be impossible for us to justify the abandon of $\delta-$meson
both conceptually and practically, while considering the highly asymmetry 
and dense nuclear system, like neutron star and relativistic heavy ion 
collision. Recent observation of neutron star like  PSR J1614-2230 with 
mass of (1.97$\pm$0.04)$M_\odot$ \cite{demo10} and the PSR J0348+0432 
with mass of (2.01$\pm$0.04)$M_\odot$ \cite{antoni13} re-open the challenge 
in the dense matter physics. 
The heavy mass of PSR J0348+0432 (M=2.01$\pm$0.04$M_\odot$) forces the nuclear 
theorists to re-think the composition and interaction inside the 
neutron star. Therefore, it is important to establish the effects of the 
$\delta$-meson and all possible interactions of other mesons for such
compact and asymmetry system.

The paper is organized as follows: In Sec. ~\ref{form}, we have outlined a brief
theoretical formalism. Here, the necessary steps of the RMF model and the
inclusion of $\delta-$meson is explained. The results and discussions are
devoted in Sec. ~\ref{resu}. Here, we have attempted to explain 
the effects of  $\delta$-meson on the nuclear matter system like 
hyperon and  proton-neutron stars. This analysis is done for both
static and rotating neutron and neutron-hyperon stars. In this
calculations, the E-RMF Lagrangian (G2 parameter set) is used 
to take care of all possible self- and crossed interactions \cite{furn97}. 
On top of the G2 Lagrangian, the $\delta-$meson interaction is added to 
take care of the isovector channel. 
The concluding remarks are given in section ~\ref{conc}.

\section{Theoretical formalism}\label{form}

From last one decade a lot of work have been done to emphasize the role 
of  $\delta-$meson on both finite and infinite nuclear matter
\cite{hofmann01,liu02,mene04,sula05}. It is seen that the contribution of 
$\delta$-meson to the symmetry energy is negative \cite{shailesh12}. To fix 
the symmetry energy around the empirical value ($\sim$30 MeV ) we need a 
large coupling constant of the $\rho-$meson $g_\rho$ value 
in the absence of the $g_\delta$. The proton and neutron effective masses 
split due to inclusion of  $\delta$-meson and consequently it affects 
the transport properties of neutron star\cite{kubis97}. The addition of
$\delta$-meson not only modify the property of 
infinite nuclear matter, but also enhances the spin-orbit 
splitting in the finite nuclei\cite{hofmann01}.  A lot of mystery 
are present in the effects of $\delta$-meson till date. The motivation
of the present paper is to study such information. It is to be noted that
both the $\rho-$ and $\delta-$mesons correspond to the isospin asymmetry, and
a careful precaution is essential while fixing the $\delta$-meson coupling
in the interaction.

The effective field theory and naturalness of the parameter are 
described in \cite{furn97,furn96,muller96,furn00,mach11}. The Lagrangian 
is consistent with underlying symmetries of the QCD. The G2 parameter is 
motivated by E-RMF theory. The terms of the Lagrangian are taken into 
account up to $4^{th}$ order in meson-baryon coupling. For the study of 
isovector channel, we have introduced the isovector-scalar $\delta$-meson. 
The baryon-meson interaction is given by \cite{bharat07}:   

\begin{eqnarray}
{\cal L}&=&\sum_B\overline{\psi}_B\left(
i\gamma^{\mu}D_{\mu}-m_B+g_{\sigma B}\sigma+g_{\delta B}\delta.\tau \right)
\psi_B \nonumber \\
&& + \frac{1}{2}\partial_{\mu}\sigma\partial_{\mu}\sigma-m_{\sigma}^2\sigma^2
\left(\frac{1}{2}+\frac{\kappa_3}{3!}\frac{g_{\sigma}\sigma}{m_B}
+\frac{\kappa_4}{4!}\frac{g_{\sigma}^2\sigma^2}{m_B^2}\right) \nonumber \\
&& - \frac{1}{4}\Omega_{\mu\nu}\Omega^{\mu\nu}+\frac{1}{2}m_{\omega}^2
\omega_{\mu}\omega^{\mu}\left(1+\eta_1\frac{g_{\sigma}\sigma}{m_B}
+\frac{\eta_2}{2}\frac{g_{\sigma}^2\sigma^2}{m_B^2}\right) \nonumber \\
&& -\frac{1}{4}R_{\mu\nu}^aR^{\mu\nu a}+\frac{1}{2}m_{\rho}^2
\rho_{\mu}^a\rho^{a\mu}\left(1+\eta_{\rho}
\frac{g_{\sigma}\sigma}{m_B} \right) \nonumber \\
&&+\frac{1}{2}\partial_{\mu}\delta.\partial_{\mu}\delta-m_{\delta}^2\delta^2
+\frac{1}{4!}\zeta_0 \left(g_{\omega}\omega_{\mu}\omega^{\mu}\right)^2
\nonumber \\&&+\sum_l\overline{\psi}_l\left(
i\gamma^{\mu}\partial_{\mu}-m_l\right)\psi_l. 
\end{eqnarray}
The co-variant derivative $D_{\mu}$ is defined as:
\begin{eqnarray}
D_{\mu}=\partial_{\mu}+ig_{\omega}\omega_{\mu}+ig_{\rho}I_3\tau^a\rho_{\mu}^a,
\end{eqnarray}
where $R_{\mu\nu}^a$ and $\Omega_{\mu\nu}$ are field tensors and
defined as follow
\begin{eqnarray}
R_{\mu\nu}^a=\partial_{\mu}\rho_{\nu}^a-\partial_{\nu}\rho_{\mu}^a
+g_{\rho}\epsilon_{abc}\rho_{\mu}^b\rho_{\nu}^c,
\end{eqnarray}
\begin{eqnarray}
\Omega_{\mu\nu}=\partial_{\mu}\omega_{\nu}-\partial_{\nu}\omega_{\mu}.
\end{eqnarray}
Here, $\sigma$, $\omega$ , $\rho$ and $\delta$ are the sigma, omega, rho and 
delta meson fields, respectively and in real calculation, we ignore the non-abelian term from the $\rho-$field. All symbols are carrying their own usual 
meaning~\cite{bharat07,sing14}.

The Lagrangian  equation for different mesons are given by \cite{bharat07}:
\begin{flushleft}
\begin{eqnarray}
m_{\sigma}^2 \left(\sigma_0+\frac{g_{\sigma}\kappa_3\sigma_0}{2m_B}
+\frac{\kappa_4 g_{\sigma}^2\sigma_0^2}{6m_B^2} \right) \sigma_0 
-\frac{1}{2}m_{\rho}^2\eta_{\rho}\frac{g_{\sigma}\rho_{03}^2}{m_B}\nonumber \\
-\frac{1}{2}m_{\omega}^2\left(\eta_1\frac{g_{\sigma}}{m_B} 
+\eta_2\frac{g_{\sigma}^2\sigma_0}{m_B^2}\right)\omega_0^2 
=\sum g_{\sigma}\rho^s_B,
\end{eqnarray}
\end{flushleft}
\begin{flushleft}
\begin{eqnarray}
m_{\omega}^2\left(1+\eta_1\frac{g_{\sigma} \sigma_0}{m_B}
+\frac{\eta_2}{2}\frac{g_{\sigma}^2\sigma_0^2}{m_B^2}\right)\omega_0 
+\frac{1}{6}\zeta_0g_{\omega}^2\omega_0^3 
=\sum g_{\omega}\rho_B,
\end{eqnarray}
\end{flushleft}
\begin{eqnarray}
 m_{\rho}^2\left(1+\eta_{\rho}\frac{g_{\sigma}\sigma_0}{m_B}\right)
=\frac{1}{2}\sum g_{\rho}\rho_{B3}
\end{eqnarray}
\begin{eqnarray}
m_{\delta}^2{\delta^3}&=&g_{\delta}^2 \rho^s_{3B}
\end{eqnarray}
with $\rho^s_{3B}=\rho^s_{p}-\rho^s_{n}$, $\rho^s_{p}$ and $\rho^s_{n}$ are
scalar densities for the proton and neutron, respectively. The total scalar 
density is expressed as the sum of the proton and neutron densities 
$\rho^s_{B} = \rho^s_{p}+\rho^s_{n}$, which is given by 
\begin{eqnarray}
\rho^s_{i}=\frac{2}{(2\pi)^3}\int_0^{k_i}\frac{M_i^*d^3k}{E_i^*} ,
i=p,n
\end{eqnarray}
and the vector (baryon) density 
\begin{eqnarray}
\rho_B=\frac{2}{(2\pi)^3}\int_0^{k_i}d^3k,
\end{eqnarray}
where, $E_i^*=(k_i^2+M_i^{*2})^{1/2}$ is the effective energy, $k_i$ is
the Fermi momentum of the baryons. $M_p^*$ and $M_n^*$ are the proton and 
neutron effective masses written as\\
\begin{eqnarray}
M_p^*=M_p-g_{\sigma}\sigma_0-g_{\delta}{\delta^3}\\
M_n^*=M_n-g_{\sigma}\sigma_0+g_{\delta}{\delta^3},
\end{eqnarray}
which is solved self-consistently. $I_3$ is the third component of
isospin projection and $B$ stands for baryon octet. The energy and pressure
density depends on the effective mass $M_B^*$ of the system, which first
needed to solve these self-consistent equations and obtained the fields
for mesons. Using the Einstein's energy-momentum tensor, the total energy 
and pressure density are given as \cite{bharat07}:

\begin{eqnarray}\label{energy}
\cal{E}&=&\sum_B\frac{2}{(2\pi)^3}\int_0^{k_B}d^3kE_B^*(k)
+\frac{1}{8}\zeta_0g_{\omega}^2\omega_0^4 \nonumber \\ 
&& + m_{\sigma}^2\sigma_0^2\left(\frac{1}{2}+\frac{\kappa_3}{3!}
\frac{g_{\sigma}\sigma_0}{m_B}+\frac{\kappa_4}{4!}
\frac{g_{\sigma}^2\sigma_0^2}{m_B^2}\right) \nonumber \\
&& + \frac{1}{2}m_{\omega}^2 \omega_0^2\left(1+\eta_1
\frac{g_{\sigma}\sigma_0}{m_B}+\frac{\eta_2}{2}
\frac{g_{\sigma}^2\sigma_0^2}{m_B^2}\right) \nonumber \\
&& + \frac{1}{2}m_{\rho}^2 \rho_{03}^2\left(1+\eta_{\rho}
\frac{g_{\sigma}\sigma_0}{m_B} \right) \nonumber \\ 
&&+\frac{1}{2}\frac{m_{\delta}^2}{g_{\delta}^{2}}(\delta^3)^2
+\sum_l\varepsilon_l,
\end{eqnarray}

and

\begin{eqnarray}\label{pressure}
\cal{P}&=&\sum_B\frac{2}{3(2\pi)^3}\int_0^{k_B}d^3kE_B^*(k)
+\frac{1}{8}\zeta_0g_{\omega}^2\omega_0^4 \nonumber \\ 
&& - m_{\sigma}^2\sigma_0^2\left(\frac{1}{2}+\frac{\kappa_3}{3!}
\frac{g_{\sigma}\sigma_0}{m_B}+\frac{\kappa_4}{4!}
\frac{g_{\sigma}^2\sigma_0^2}{m_B^2}\right) \nonumber \\
&& + \frac{1}{2}m_{\omega}^2 \omega_0^2\left(1+\eta_1
\frac{g_{\sigma}\sigma_0}{m_B}+\frac{\eta_2}{2}
\frac{g_{\sigma}^2\sigma_0^2}{m_B^2}\right) \nonumber \\
&& + \frac{1}{2}m_{\rho}^2 \rho_{03}^2\left(1+\eta_{\rho}
\frac{g_{\sigma}\sigma_0}{m_B} \right) \nonumber \\
&&-\frac{1}{2}\frac{m_{\delta}^2}{g_{\delta}^{2}}(\delta^3)^2+\sum_l P_l,
\end{eqnarray}
where $P_l$ and $\varepsilon_l$ are lepton's pressure and energy 
density, respectively.

\section{Results and discussions}\label{resu}

Before going to the discussions of our results, we give 
a brief description of the parameter fitting procedure for 
g$_\rho$ and g$_\delta$ in subsection ~\ref{resua}. 
Then the results on hyperon star along with the neutron star structures 
both for static and rotating cases under $\beta-$equilibrium condition 
are discussed in the subsequent subsections ~\ref{resub}, ~\ref{resuc},~\ref{resud}, ~\ref{resue}  and ~\ref{resuf}.

\subsection{Parameter Fitting}\label{resua}

It is important to fix the $g_\delta$  value to see the effects of the 
$\delta$-meson. The isovector channels in  RMF theory come to exist 
through both the $\rho$ and $\delta$ mesons couplings. 
While considering the effects of the $\delta$-meson, we have to 
take the $\rho $-meson into account. Since both the isovector channels are
related to isospin, one can not optimize the $g_\delta$ coupling
independently. Here, we have followed a more reliable procedure by fixing
the symmetry energy $E_s$ by adjusting simultaneously different values of 
$g_\rho$ and $g_\delta$ value\cite{kubis97}. As it is mentioned earlier, 
we have added $g_\delta$ on top of the G2 parameter set. Thus, the symmetry 
energy of G2 parameter is $E_s = 36.4 $ MeV is kept constant at the time of
re-shuffling $g_\rho$ and $g_\delta$. The G2 parameters and the $g_{\delta}$
and $g_{\delta}$ combinations are displayed in Table ~\ref{tab3}. The
nuclear matter properties are also listed in the table.
\begin{figure}[ht]
\includegraphics[width=1.1\columnwidth]{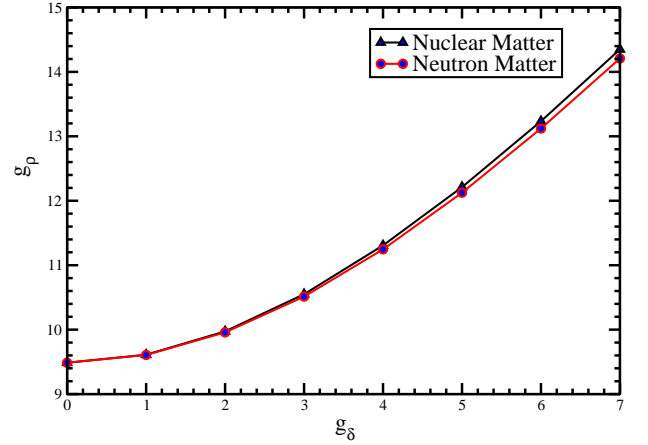}
\caption{(Color online)Variation of $g_\rho$ and $g_\delta$ at
a constant value of symmetry energy $E_s = 36.4 $ MeV for both nuclear and neutron 
matter.
}
\label{fig1}
\end{figure}

\begin{table*}
\hspace{0.1 cm}
\caption{The force parameters for G2 set are given in the upper panel of 
the Table. The nuclear matter saturation properties are given in the middle
panel and various $g_{\rho}$ and $g_{\delta}$ combinations are given in the
lower panel, keeping symmetry energy E$_{s}$ = 36.4 MeV fixed.
}
\renewcommand{\tabcolsep}{0.01 cm}
\renewcommand{\arraystretch}{1.}
{\begin{tabular}{|c|c|c|c|c|c|c|c|c|c|c|c|c|c|c|}
\hline
\hline
$m_{n}$ = 939.0 MeV& $m_{\sigma}$ = 520.206 MeV&$m_{\omega}$ = 782.0 MeV &
$m_{\rho}$ = 770.0 MeV& $m_{\delta}$ = 980.0 MeV &$\Lambda$= 0.0&
$\zeta_{0}$ = 2.642&$\eta_{\rho}$ = 0.39&\\ 
$g_{\sigma}$ = 10.5088&$g_{\omega}$ = 12.7864& $g_{\rho}$ = 9.5108&$g_{\delta}$ = 0.0&$k_{3}$
 = 3.2376&$k_{4}$ = 0.6939&$\eta_{1}$ = 0.65& $\eta_{2}$ = 0.11& \\
\hline
$\rho_{0}$ = 0.153$fm^{-3}$&E/A = -16.07MeV& $K_{\infty}$ = 215 MeV  
&$E_{s}$ = 36.4 MeV& $m^*_{n}/m_{n}$ = 0.664 &&&&  \\
\hline
\hline
$(g_{\rho}, g_{\delta})$&(9.510, 0.0)&(9.612, 1.0)&(9.973, 2.0)&(10.550, 3.0)&(11.307, 4.0)&
(12.212, 5.0)&   
(13.234, 6.0) &  
(14.349, 7.0)\\
\hline
\end{tabular}\label{tab3}}
\end{table*}
For a particular value of $E_s = 36.4 $ MeV, the variation of $g_\rho$ and $g_\delta$ are plotted in Fig.~\ref{fig1}. From Fig.~\ref{fig1},  it is clear that 
as the $g_\delta$ increases 
the $g_\rho$ value also increases, almost linearly, to fix the symmetry energy
unchanged. This implies that $\rho$ and $\delta$-mesons have opposite effect on 
$E_s$ contribution, i.e.,  the $\delta$-meson has negative contribution 
of the symmetry energy contrary to the positive contribution of 
$\rho$-meson. 

\subsection{Fields of $\sigma, \omega, \rho$ and $\delta$ mesons}\label{resub}

\begin{figure}[ht]
\includegraphics[width=1.1\columnwidth]{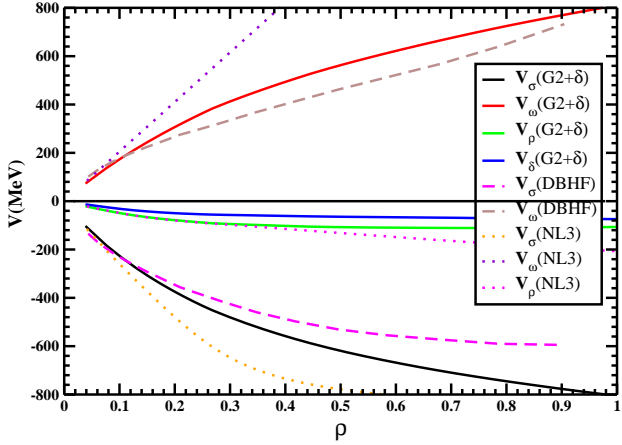}
\caption{(Color online)Various meson fields are obtained from the 
RMF theory with $G2+g_{\delta}$ and NL3 parameter sets. The $\sigma-$ 
($V_\sigma$) and $\omega-$ ($V_\omega$) fields are compared with the results 
of DBHF theory \cite{brok90}.}
\label{fig2}
\end{figure}

The fields of the meson play a crucial role to construct the nuclear 
potential, which is the deciding factor for all type of calculations in
the relativistic mean field model. In Fig.~\ref{fig2}, we have plotted 
various meson fields included in the present calculations, such as 
$\sigma$, $\omega$, $\rho$ and $\delta$ with $g_{\delta}$ on top of 
G2 parameter set $(G2+g_{\delta})$. 
It is obvious that $V_\sigma$ and $V_\omega$ are opposite to each
other, which is also reflected in the figure. This means, the positive
value of $V_{\omega}$ gives a strong repulsion, which is compensated by the
strongly attractive potential of the $\sigma-$meson field $V_\sigma$.
The nature of the curves for $V_\sigma$ and $V_\omega$ are almost similar 
except the sign. The magnitude of $V_\sigma$ and $V_\omega$ looks
almost equal. However, in real (it is not clearly visible in the curve,
because of the scale), the value of $V_\sigma$ is slightly larger
than $V_\omega$, which keeps the overall nuclear potential strongly attractive.
The attractive $V_\sigma$ and repulsive $V_\omega$ potentials  combinely 
give the saturation properties of the nuclear force. It is worthy to mention
that the contributions of self-interaction terms are taken care
both in $V_\sigma$ and $V_\omega$, which are the key quantities to solve the
Coester band problem \cite{bidhu14} and the explanation of quark-gluon-plasma 
(QGP) formation within the relativistic mean field formalism \cite{subrat15}.
The self-interaction of the $\sigma-$meson gives a repulsive force at long
range part of the nuclear potential, which is equivalent to the 3-body 
interaction and responsible for the saturation properties of nuclear force.
The calculated results of $V_\sigma$ and $V_\omega$ are compared with the 
results obtained from DBHF theory with Bonn-A potential\cite{brok90}.

Fig.~\ref{fig2} clearly shows that in the low density region 
(density $\rho$ $\sim 2\rho_0$) both RMF and DBHF theories well matched. 
But as it increases beyond density $\rho$ $\sim 2\rho_0$ both the calculations 
deviate from each other. 
The possible reason may be the fitting of parameters in Bonn-A potential 
is up to $2-3$ times of saturation density $\rho_0$, beyond that the DBHF
data are simple extrapolation of the DBHF theory.
The contribution of both $\rho-$ and $\delta-$ mesons correspond to the 
isovector channel. The $\delta-$meson gives different effective masses 
for proton and neutron, because of their opposite iso-spin of the third
component. The nuclear potential generated by the $\rho-$ and $\delta-$mesons
are also shown in  Fig~\ref{fig2}. We noticed that although their contributions
are small, but non-negligible. These non-zero values of $V_{\rho}$ and 
$V_{\delta}$ to the nuclear potential has a larger consequence, mostly
in compact dense object like neutron or hyperon stars, which will be
discussed later in this paper.

\subsection{Energy per particle and pressure density}\label{resuc}
 
\begin{figure}[ht]
\includegraphics[width=1.1\columnwidth]{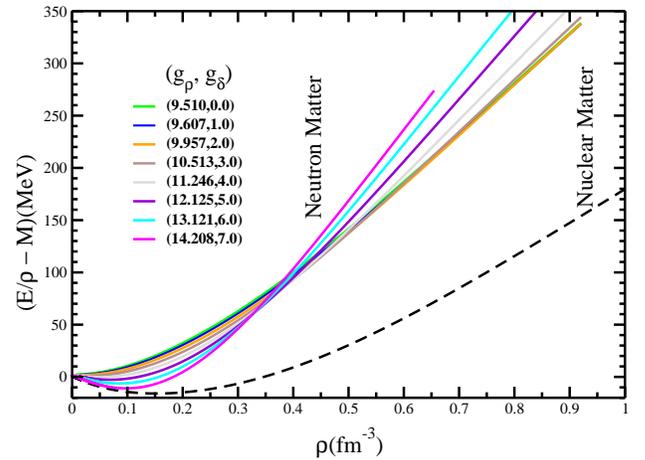}
\caption{(Color online)  Variation of binding energy per particle
with density at various $g_\rho$ and $g_\delta$. }
\label{fig3}
\end{figure}

\begin{figure}[ht]
\includegraphics[width=1.1\columnwidth]{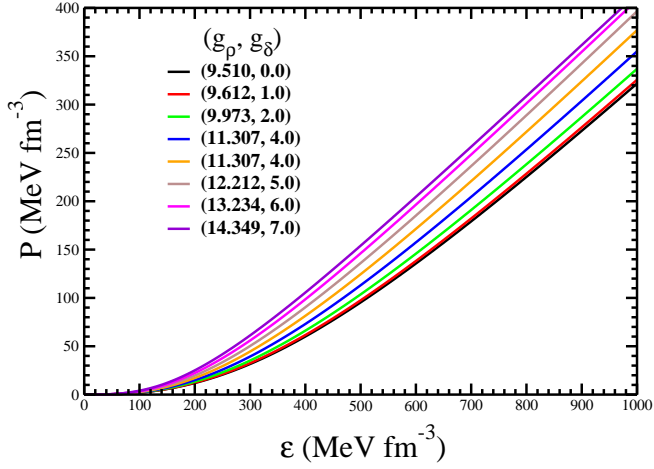}
\caption{(Color online)  Variation of EOS for different values of $g_\rho$ 
 and $g_\delta$. }
\label{fig4}
\end{figure}

The energy density and pressure density are known as equations of states (EOS). 
These quantities are the key ingredients to describe the structure of 
neutron/hyperon stars. To see the sensitivity of the EOS, we have plotted  
energy per particle ($E/\rho -M$)  as a function of density for pure
neutron matter in  Fig~\ref{fig3} and pressure density as a function of 
energy density in Fig~\ref{fig4}. Each curve corresponds to a particular 
combination of $g_\delta$ and $g_\rho$, which reproduce the symmetry 
energy $E_s = 36.4 $ MeV without destabilizing other parameters of G2 set.
The green line represents for $g_\delta =0$, i.e., with pure G2 parameter set. 
Both the binding energy per particle as well as the pressure density increase
with the value of $g_\delta$.
This process continue till the value of $g_\delta$ reaches, at which 
$E/\rho -M$ equals the nuclear matter binding energy per particle. An
unphysical situation arises beyond this value of $g_\delta$, because the
binding energy of the neutron matter will be greater than $E/\rho -M$ for
symmetric nuclear matter.
In the case of G2+$\delta$ parametrization, this limiting 
value of $g_{\delta}$ reaches at $g_\delta$= 0.7, after which 
we do not get a convergence solution in our calculations.

\subsection{Stellar properties of static and rotating neutron stars}\label{resud}

The $\beta$-equilibrium and  charge neutrality  
are two important conditions to justify the structural composition 
of the neutron/hyperon stars. Both these conditions force the stars to 
have $\sim$90$\%$ of neutron and $\sim$10$\%$ proton. With the inclusion of  
baryons, the $\beta-$equilibrium conditions between chemical potentials
 for different particles: 
\begin{eqnarray}\label{beta}
\mu_p = \mu_{\Sigma^+}=\mu_n-\mu_{e} \nonumber\\
\mu_n=\mu_{\Sigma^0}=\mu_{\Xi^0}=\mu_{n}\nonumber \\
\mu_{\Sigma^-}=\mu_{\Xi^-}=\mu_n+\mu_{e}\nonumber \\
\mu_{\mu}=\mu_{e} \nonumber\\
\end{eqnarray}
and the charge neutrality condition is satisfy by  
\begin{eqnarray}\label{charge}
n_p+n_{\Sigma^+}=n_e+n_{\mu^-}+n_{\Sigma^-}+n_{\Xi^-}
\end{eqnarray}

To calculate the mass and radius profile of the static (non-rotating), but 
spherical neutron star, we solve the general relativity
Tolmann-Oppenheimer-Volkov (TOV)\cite{tolm39} equations which are written as: 
\begin{eqnarray}
\frac{d P(r)}{d r}=-\frac{G}{c^2}\frac{[{\cal E}(r)+P(r)][M(r)+\frac{4\pi r^3 P(r)}{c^2}]}{r^2(1-\frac{2GM(r)}{c^2 r})}
\end{eqnarray} 
and
\begin{eqnarray}
\frac{d M(r)}{d r}=\frac{4\pi r^2 {\cal E}(r)}{c^2},
\end{eqnarray}
with G as the gravitational constant, $\cal E$$(r)$ as the energy density, 
$P(r)$ as the pressure density and $M(r)$ as the gravitational mass inside 
radius $r$. We have used c=1. For a given EOS, these equations can be 
integrated from the origin as an initial value problem for a given choice 
of the central density $\cal E$$_c$($r$). The value of r( = R) at which the 
pressure vanishes defines the surface of the star. 
In order to understand the effect of $\delta-$meson coupling on neutron
star structure, we must also look, what happens to massive objects as they 
rotate and how this affects the space-time around them. 
\begin{figure}[ht]
\includegraphics[width=1.1\columnwidth]{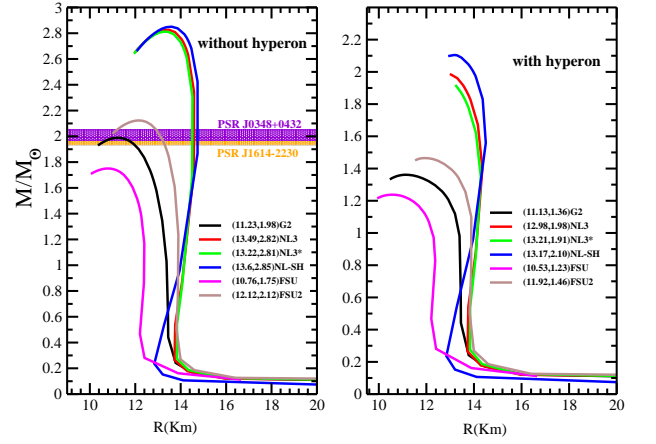}
\caption{(Color online) The mass-radius profile for static star with different 
parametrizations like G2\cite{furn97}, NL3\cite{lala97}, NL3*\cite{lala09}, 
NL-SH\cite{sharma93}, FSU\cite{pieka05}and FSU2\cite{fsu2}. (a) The left panel is for 
proton-neutron star and (b) the right panel is for the hyperon star.
}
\label{fig5}
\end{figure}
For this, we use the code written by Stergioulas\cite{ster95} based on Komastu, 
Eriguchi, and Hachisu (KEH) method (fast rotation)\cite{kom89,koma89} to 
construct mass-radius of the uniform rotating star. One should note that the
maximum mass of static star is less than the rotating stars. Because, 
when the massive objects rotate they flatten at their poles. The forces of 
rotation, namely the effective centrifugal force,  pulls the mass farthest 
from the center further out, creating the equatorial bulge. This pull away 
from the center will, in part, counteract  gravity, allowing the star to be 
able to support more mass than its  non-rotating star.

We know that the core of neutron stars contain hyperons a very high density
($\sim$7-8 $\rho{_0}$) matter. As it is mentioned before, with the presence of 
baryons, the EOS becomes softer and stellar properties will be changed. The 
maximum mass of hyperon star decreases about 10-20$\%$ depending on the choice 
of the meson-hyperon coupling constants. The 
hyperon couplings are expressed as the ratio between the meson-hyperon and 
meson-nucleon couplings as:
\begin{eqnarray}
\chi_{\sigma} = \frac {g_{Y\sigma}}{g_{N\sigma}}, \chi_{\omega} = \frac {g_{Y\omega}}{g_{N\omega}},\chi_{\rho} = \frac {g_{Y\rho}}{g_{N\rho}}, \chi_{\delta} = \frac {g_{Y\delta}}{g_{N\delta}}.
\end{eqnarray}
In the present calculations, we have taken $\chi_{\sigma} = \chi_{\rho} = 
\chi_{\delta}$ = 0.6104 and $ \chi_{\omega}$ = 0.6666\cite{glen91}. One can
find similar calculations for stellar mass in Refs. \cite{glen20,weis12,lope14}. 
Now we present the star properties like mass and radius in Figs. 
~\ref{fig5}, ~\ref{fig6} and ~\ref{fig7}. In Fig.~\ref{fig5} we  plotted 
the mass-radius profile for the proton-neutron star as well as 
for the hyperon star using a wide variation of parameter sets starting
from the old parameter like NL-SH\cite{sharma93} to the new set of
FSU2 \cite{fsu2}. The mass-radius profile varies 
to a great extend over the choice of the parameter. For example, in FSU
parameter set \cite{pieka05}, the maximum possible mass of the proton-neutron 
star is $\sim$ 1.75 $M_\odot$, while the maximum possible mass for the 
NL3 set \cite{lala97} is $\sim$ 2.8 $M_\odot$. 
These results are shown in the left panel of the Fig.~\ref{fig5}, while 
right panel show same things for the hyperon star. 

\subsection{Effects of $\delta-$meson on static and rotating stars}\label{resue}

\begin{figure}[ht]
\includegraphics[width=1.1\columnwidth]{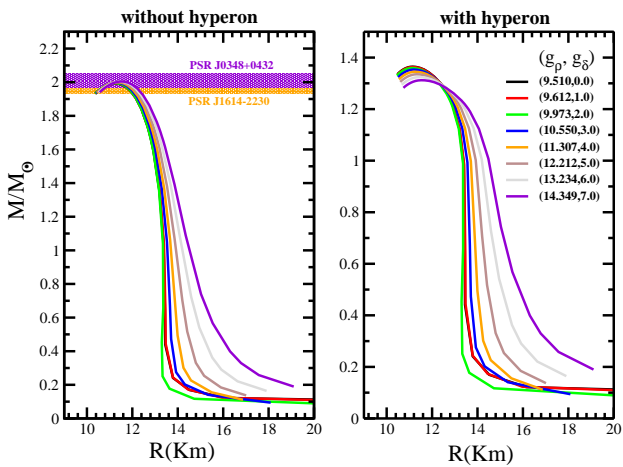}
\caption{(Color online) The mass-radius profile of the static proton-neutron 
and hyperon stars with various combination of $g_{\delta}$ and $g_{\rho}$ in 
G2+$\delta$. (a) The left panel is for proton-neutron star and (b) the right 
panel is for the hyperon star.}
\label{fig6}
\end{figure}

\begin{figure}[ht]
\includegraphics[width=1.1\columnwidth]{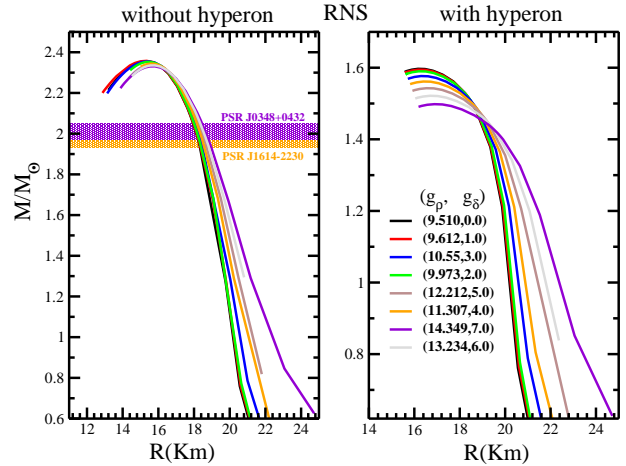}
\caption{(Color online) Same as Fig~\ref{fig6}, but for rotating stars.}
\label{fig7}
\end{figure}

The main aim of  this paper is to understand the effects of $\delta$-meson 
on  neutron stars both with and without hyperons. 
Figs.~\ref{fig6} and \ref{fig7} represent the mass-radius profiles for 
non-rotating stars taking into account the presence of with and without 
hyperons.  These profiles are shown for various combinations of $g_{\rho}$ 
and $g_{\delta}$, which we have obtained by fitting the symmetry energy
$E_s$ of pure nuclear matter.  
Analyzing the graphs, we notice a slight change in the maximum mass with 
g$_\delta$ value. That means, the mass of the star goes on decreasing with
an increase value of the $\delta$-meson coupling. A further inspection of the
results reveals that, although the  $\delta$-meson coupling has a 
nominal effects on the maximum mass of the stars, we get an asymptotic 
increase in the radius. This asymptotic nature of the curves is more
prominent in presence of hyperons inside the stars. Similar phenomena are
also observed in case of rotating stars.  

\subsection{Effects of $\delta-$meson on baryon production}\label{resuf}
Finally, we want to see the effects of $\delta-$meson coupling on the 
particle production for the whole baryonic family at various densities
in nuclear matter system.  
\begin{figure}[ht]
\includegraphics[width=1.1\columnwidth]{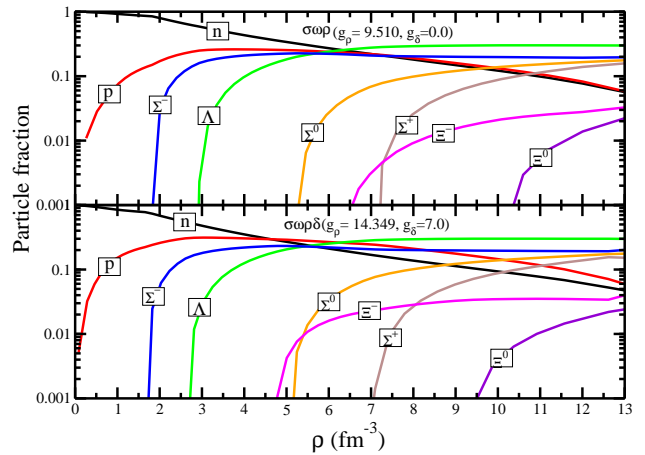}
\caption{(Color online) Yield of strange particles as a function of density. 
The upper panel is with G2 parameter set (without taking $\delta-$meson 
coupling) and the lower panel is with $\delta-$meson coupling. }
\label{fig8}
\end{figure}
The Fermi energy of both proton and neutron increases with density 
for their Fermionic nature. After a certain density, the Fermi energy of the
nucleon exceeded the rest mass energy of the nucleon ($\sim$1000 MeV), 
and strange particles ($\Sigma, \Lambda,\Xi$) are produced. As a result, the
equations of state of the star becomes soft and gives a smaller star mass 
compare to the neutron star containing only protons, neutrons and electrons. 
The decrease in star mass in the presence of whole baryon octet can be
understood from the analysis of Fig.~\ref{fig8}. From the figure it is clear 
that $\delta$-meson has a great impact on the production of hyperons. The
inclusion of $\delta-$meson accelerate the strange particle production. 
For example, the evolution of $\Sigma^-$ takes place at density 
$\rho=1.75\rho_0$ in absence of $\delta-$meson. However, it produces at
$\rho=1.67\rho_0$ when $\delta-$meson is there in the system. Similarly,
analyzing the evolution of other baryons, we notice that although the
early production of baryons in the presence of $\delta-$meson is not 
proportionate to each other, in each case the yield is faster. A significant
shifting towards lower density is maximum for heaviest hyperon ($\Xi^0$) and
minimum for nucleon (see Fig. ~\ref{fig8}). For example,  $\Xi^{-}$ evolves
at $\rho_{B}$ = 6.5 $\rho_{0}$ for a non-$\delta$ system and $\rho_{B}$$\sim$5.0 $\rho_{0}$ for medium when $\delta-$meson is included. 
Thus, the $\delta-$coupling has a sizable impact on the production of 
hyperons like  $\Xi^{-}, \Xi^{0}, \Sigma^{+}$.

\section{Summary and Conclusions}\label{conc}

In summary, using the effective field theory approach, we discussed the effect
of isovector scalar meson on hyperon star. Inclusion of $\delta$-meson with
G2 parameter set, we have investigated the static and rotating stellar 
properties of neutron star with hyperons. We fitted the parameter and see the
variation of g$_\rho$ and g$_\delta$ at a constant symmetry energy for 
both the nuclear and neutron matter.  
With the help of G2+$\delta$ model, for static and rotating stars without 
hyperon, core we get the maximum mass of $\sim$2$M_\odot$ and 
$\sim$2.4$M_\odot$, respectively. This prediction of masses is in agreement 
with the recent observation of $M\sim$$2M_\odot$ of the stars. However, with 
hyperon core the maximum mass obtained are $\sim$1.4$M_\odot$ and 
$\sim$1.6$M_\odot$ for static and rotating hyperon stars, respectively. 
In addition, we have also calculated the production of whole
baryon octet with variation in density. We find that the particle fraction 
changes a lot in presence of $\delta-$meson coupling. When there is
$\delta-$meson in the system the evolution of baryons are faster compare
to a non-$\delta$ system. This effect is significant for heavier masses and
minimum for lighter baryon. Hence, one can conclude that the yield of 
baryon/hyperons depends very much on the mesons couplings.

\end{document}